\documentclass[aps,prx,reprint,showpacs,floatfix,superscriptaddress]{revtex4-2}
\usepackage{xcolor}
\usepackage{microtype}
\usepackage[bookmarks=true,
bookmarksnumbered=false, 
bookmarksopen=false, 
colorlinks=true,
linkcolor=blue]{hyperref}
\definecolor{webblue}{rgb}{0, 0, 0.5} 
\hypersetup{colorlinks=true,urlcolor=blue,citecolor=blue}
\usepackage{graphicx}
\usepackage{subfigure}
\usepackage{url}
\usepackage{hyperref}
\usepackage{amsmath}
\usepackage[capitalize]{cleveref}
\usepackage{braket}
\usepackage{outlines}
\usepackage{enumitem}
\usepackage{soul}
\usepackage{color}
\usepackage{bm} 

\usepackage[utf8]{inputenc}
\usepackage{multirow}
\usepackage{amssymb}

\usepackage{relsize}

  \usepackage{footnote}

\usepackage{soul}

\usepackage{pgfplots}
\usepackage{ifthen}

\begin{document}

\title{Exciton condensation in biased bilayer graphene}

\author{Harley D. Scammell}
\affiliation{School of Physics, the University of New South Wales, Sydney, NSW, 2052, Australia}

\author{Oleg P. Sushkov}
\affiliation{School of Physics, the University of New South Wales, Sydney, NSW, 2052, Australia}

\date{\today}

\begin{abstract}
  We consider suspended bilayer graphene under applied perpendicular electric bias field
 that is known to generate a single particle gap $2\Delta$ and a related
 electric polarization ${\cal P}$. We argue that
 the bias also drives a quantum phase transition from band insulator to superfluid exciton condensate. The transition occurs when the exciton binding energy exceeds the band gap $2\Delta$. 
We predict the critical bias (converted to band gap), $\Delta_c\approx 60$ meV, below which the excitons condense. 
The critical temperature, $T_c(\Delta)$, is maximum at $\Delta \approx 25$ meV, $T_c^\text{max}\approx 115$ K,  decreasing significantly at smaller  $\Delta$ due to thermal screening.
Entering the condensate phase, the superfluid transition is accompanied by a cusp in the  electric polarization ${\cal P}(\Delta)$ at $\Delta\to\Delta_c$, which provides a striking testable signature. Additionally, we find that the condensate prefers to form a pair density wave.

\end{abstract}

\maketitle


{\it Introduction ---}
Excitonic condensates in two-dimensional (2D) materials promise novel superfluid \cite{Lozovik1975, pogrebinskii1977mutual, blatt1962bose, kellogg2004vanishing, su2008make} or topological \cite{Wang2019, Varsano2020, Perfetto2020, Sun2021, Liu2021topo,  scammell2022chiral} properties, and have thereby attracted considerable theoretical and experimental attention. Moreover, due to these novel transport behaviours, excitonic condensates promise a route to future technological advancements. 
So far, however, experimental realisations of the desired condensate have proven problematic. 

Theoretically it is convenient to classify exciton condensates (also known as exciton insulators) by the nature of the corresponding interaction-driven quantum phase transition. {\it Class I} corresponds to the
(semi)metal-to-exciton condensate phase transition \cite{KeldyshKopaev1965, RiceKohn1967, Halperin1968}. In this case there are simultaneous Fermi surfaces of electrons and holes. If the Fermi surfaces are identical, an arbitrarily weak attraction between electrons and holes leads to condensation. This situation is analogous to BCS superconductivity. {\it Class II} corresponds to the band insulator-to-exciton condensate phase transition. In this case there is no Fermi surface and the interaction must exceed a critical value to generate the condensate.

Graphene layers, with their very near particle-hole symmetry, have provided a hunting ground for exciton condensation. Previous theoretical considerations include bilayer graphene (BLG) --- unbiased with AB stacking \cite{Nandkishore2010, Song2012, Apinyan2016} or biased with AA stacking \cite{Akzyanov2014, Apinyan2021}.
There have been no corresponding experimental detections of condensation.

Another set of proposals are graphene double layers \cite{Min2008, Lozovik2008,  Zhang2008, Kharitonov2008, Kharitonov2010} or double bilayers \cite{Perali2013, Su2017, EfimkinBurg2020, Neilson2019}, separated by a dielectric. Theoretical predictions for the Berezinskii–Kosterlitz–Thouless (BKT) transition temperature for such systems vary significantly from room temperature in Ref. \cite{Min2008} to 1 mK in Refs. \cite{Kharitonov2008, Kharitonov2010} --- the key difference arises due to the inclusion \cite{Kharitonov2008, Kharitonov2010}  or exclusion  \cite{Min2008}  of Coulomb screening. 
Experimentally, there is one recent indirect indication of possible zero-magnetic field exciton
condensation in double-bilayer graphene with WSe$_2$ spacer \cite{Burg2018} and also in an InAs/GaSb bilayer \cite{Du2017}, with both scenarios belonging to Class I.
On the other hand, there have been several experimental reports of exciton condensation in quantum Hall regime in strong magnetic field for double-layer graphene  \cite{Liu2017, LiDean2017, LiDean2019} or other double layer systems \cite{Kellogg2004, TutucHuse2004, Wiersma2004}; such excitonic pairing occurs between different Landau levels.

Most, if not all, previous studies of exciton condensation have been aimed at class I. However, class I is necessarily a many-body problem and progress often requires uncontrolled approximations. Here, instead, we consider condensation in class II. A striking technical advantage is that we only need to consider a two-body problem. In class II the condition of exciton condensation is the equality of the exciton binding energy $\epsilon_b$ to the single particle band gap, $2\Delta$. 
The present work is focused on biased BLG. 
Unbiased BLG is a semimetal, however, application of an electric bias (perpendicular electric field) opens a single particle band gap $2\Delta$ \cite{McCann2013} -- placing it within class II. The bias is created by symmetric metallic gates above and below the plane. 

A careful treatment of the screened attractive electron-hole Coulomb interaction is essential to make quantitative predictions of the binding energy and hence of the condensation transition. To this end, we account for three sources of screening: (i) Screening by metallic gates placed a distance $d$ above and below the BLG plane;
 (ii) dielectric screening due to a material between BLG and the gates; and 
 (iii) BLG self-screening and retardation thereof, as captured via the Random Phase Approximation (RPA). We find that all three sources of screening significantly influence, i.e. reduce, the condensation critical temperature. To this end, we propose idealised experimental setups to allow to maximise the critical temperature. A key development in this work is our treatment of the retardation of the self-screening, i.e. retardation of the screened Coulomb potential. 
 
 The key development of this work is that we are able to: use reliable two-body techniques to describe excitonic bounds states in a band insulator; account for environmental sources of screening; treat dynamically screening/retardation. This combination affords a reliable prediction of the critical temperature, and how it may be optimised with respect to system parameters. 
Entering the exciton condensate phase, we lose this quantitative control over the problem, and instead resort to mean-field theory. We note that the exciton condensate is a neutral superfluid, and that experimentally distinguishing between this and the band insulator phase is subtle. Using our mean-field description, we establish two key experimental signatures of the excitonic condensate. 

 

\begin{figure}[t!]
  \includegraphics[width=0.375\textwidth]{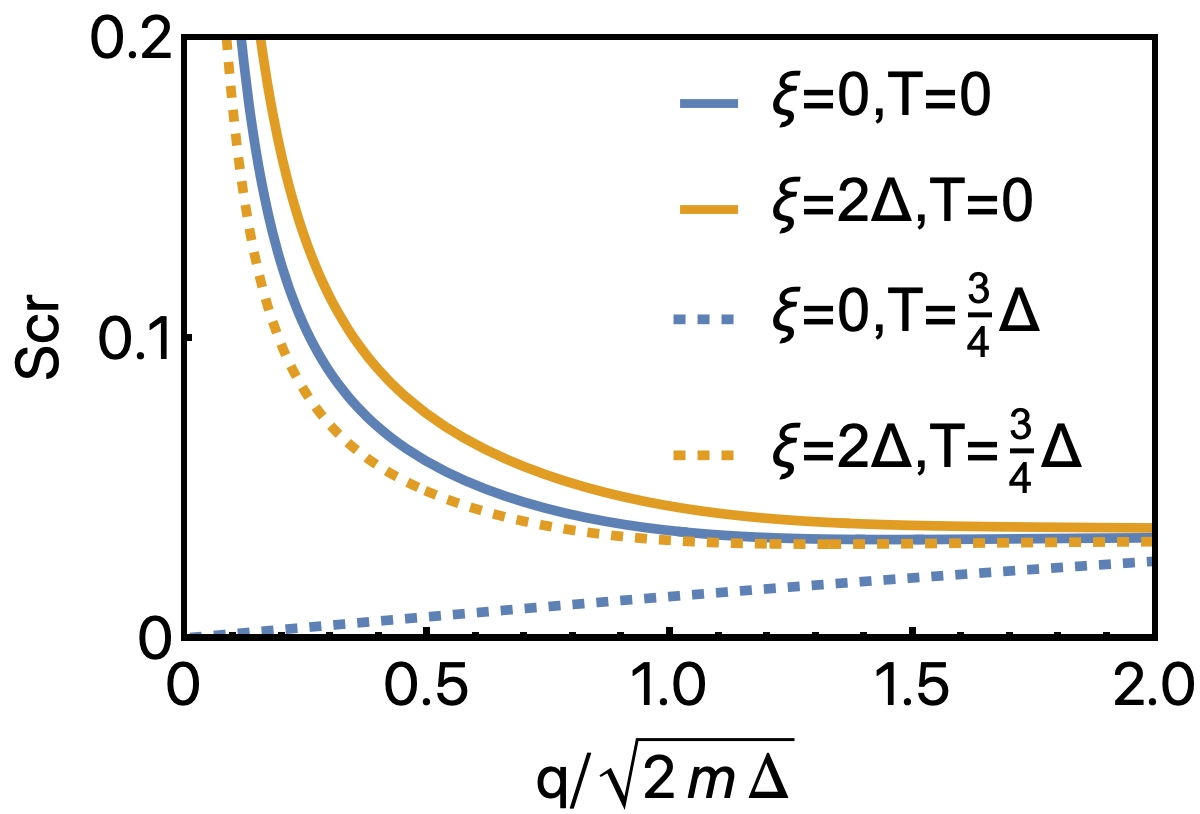}
    \caption{Screening factor vs momentum for $\Delta=10$meV,
d=1000nm, $\epsilon_r=1$.
              The factor is presented for two imaginary frequencies,
              $\xi=0$, $\xi=2\Delta$, and for two temperatures, $T=0$, $T=3\Delta/4= 87$ K.
     }\label{scr}
\end{figure}

\

{\it Methods} --- We consider biased bilayer graphene for which the low-energy single particle Hamiltonian can be reduced to \cite{Falko2006}
\begin{eqnarray}
\label{H1}
  H_{0}=\left(
  \begin{array}{c c}
    \Delta -\mu & -\frac{p_-^2}{2m}\\
 -\frac{p_+^2}{2m} &-\Delta -\mu
  \end{array}
  \right).
\end{eqnarray}
$H_{0}$ is written in terms of $\{A_1, B_2\}$ orbitals, with $A,B$ referring to graphene sublattice and subscripts $1,2$ referring to layers. 
Here $p_\pm = \tau p_x \pm i p_y$, $\bm p$ is the in-plane momentum,
$\tau=\pm 1$ the valley quantum number, $m\approx 0.032m_e$ the effective mass,
and $\Delta$ is proportional to the bias electric field, $\Delta \propto E$,
\cite{Zhang2009}.
The chemical potential $\mu$ is set to zero (half-filling) for the rest of
this work.
There are corrections to (\ref{H1}) related to electron-hole asymmetry, 
trigonal warping, etc. However, influence of all these corrections on
the exciton is negligible at $\Delta < 60$meV,
see Ref.\cite{ScammellSushkov2022}, so here we disregard the corrections.
Hence electron and hole dispersions, denoted $\omega_{\bm p}^{(\pm)}$, are symmetric about $\mu=0$, i.e. $\omega_{\bm p}^{(\pm)}=\pm\omega_{\bm p}$, with 
$\omega_{\bm p}=\sqrt{\Delta^2+p^4/(4m^2)}$.
Application of the bias induces electric polarisation
along the field. If $a\approx 0.3$nm is the separation between the planes
and $e$ is the electron charge, the electric dipole moment per unit area is
$|e|a{\cal P}_\Delta$, with the layer polarization ${\cal P}_\Delta$, due to the bias $\propto\Delta$, given by
\begin{eqnarray}
  \label{P}
        {\cal P}_{\Delta}=4\int \frac{d^2p}{(2\pi)^2}(|\beta^{(-)}_{\bm p}|^2-|\alpha^{(-)}_{\bm p}|)=
        \frac{2m\Delta}{\pi}\ln\frac{\Lambda}{|\Delta|}.
\end{eqnarray}
The factor 4 is due to the spin and valley degeneracy, $\alpha^{(-)}_{\bm p,\tau}$
and $\beta^{(-)}_{\bm p,\tau}$ are upper and lower
components of the negative energy eigenfunction (for a given valley), and $\Lambda$
is the ultraviolet energy cutoff. We confirm via a direct calculation that including the next two orbitals, i.e. $\{A_2, B_1\}$, naturally cuts off of the UV divergence; the choice $\Lambda = 0.5$ eV in (\ref{P}) is consistent with the direct four-orbital calculation.

\begin{figure}[t!]
  \includegraphics[width=0.45\textwidth]{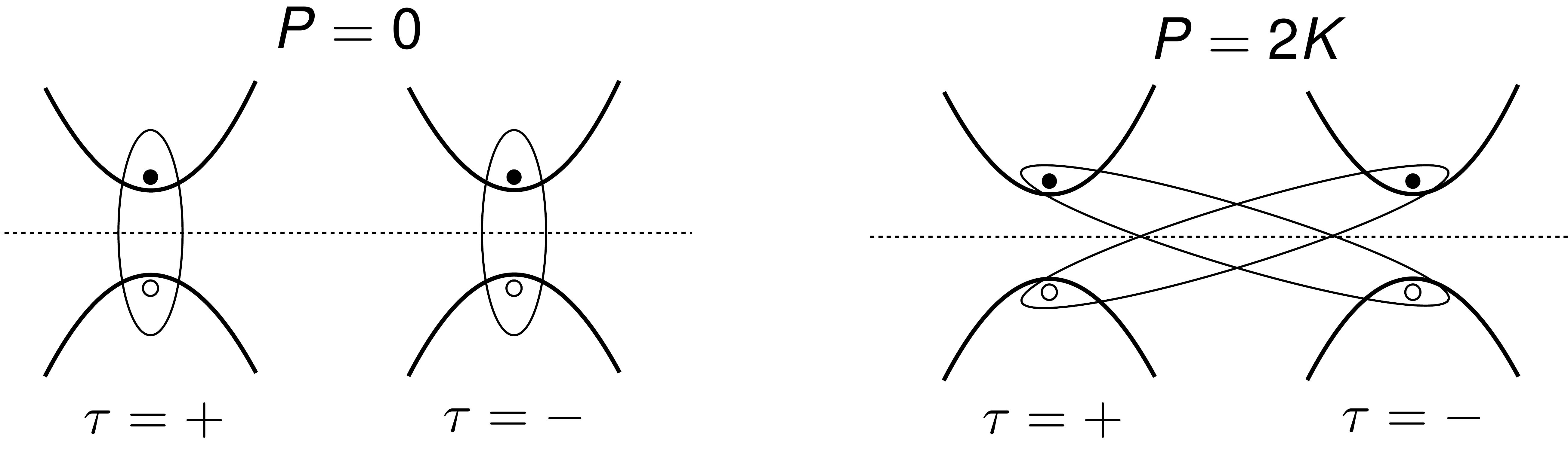}
    \caption{Inter- and intra-valley excitons: intravalley have zero total momentum $\bm P=\bm 0$, while intervalley have $\bm P=2\bm K$, with $\bm K$ the valley momentum.}\label{valley}
\end{figure}

Attraction between electron and hole is due to the screened Coulomb interaction
\begin{align}
\label{Int}
V_{\bm q, i\xi}=-\frac{2\pi e^2}{\epsilon_r q/\Upsilon_q-2\pi e^2\Pi({\bm q},i\xi, T)}
\end{align}
Here $\bm q$ is the momentum transfer, $\xi$ is the imaginary frequency transfer,
and $T$ is temperature. Other parameters are:
(i) $\Upsilon_p=\tanh(p d)$ accounts for the metallic gate screening,
at a distance $d$ above and below the BLG plane;  
(ii) $\epsilon_r$ is the dielectric constant of the
substrate/superstrate material between BLG and gates; and 
(iii) $\Pi({\bm q},i\xi, T)$ is the polarisation operator of BLG, with details presented in the Supplement \cite{Supp}.
To demonstrate key features of (\ref{Int}), we introduce a screening factor -- defined as the
ratio of the screened interaction (\ref{Int})
to the bare Coulomb interaction, $-2\pi e^2/(\epsilon_r q)$. The screening factor provides a measure of the effectiveness of screening. 
Fig. \ref{scr} demonstrates that the screening very strongly depends on
frequency and on temperature. Notably, thermally excited electrons practically fully
screen the static interaction, i.e. the screening factor becomes vanishingly small.

\begin{figure*}[t!]
    \includegraphics[width=0.375\textwidth]{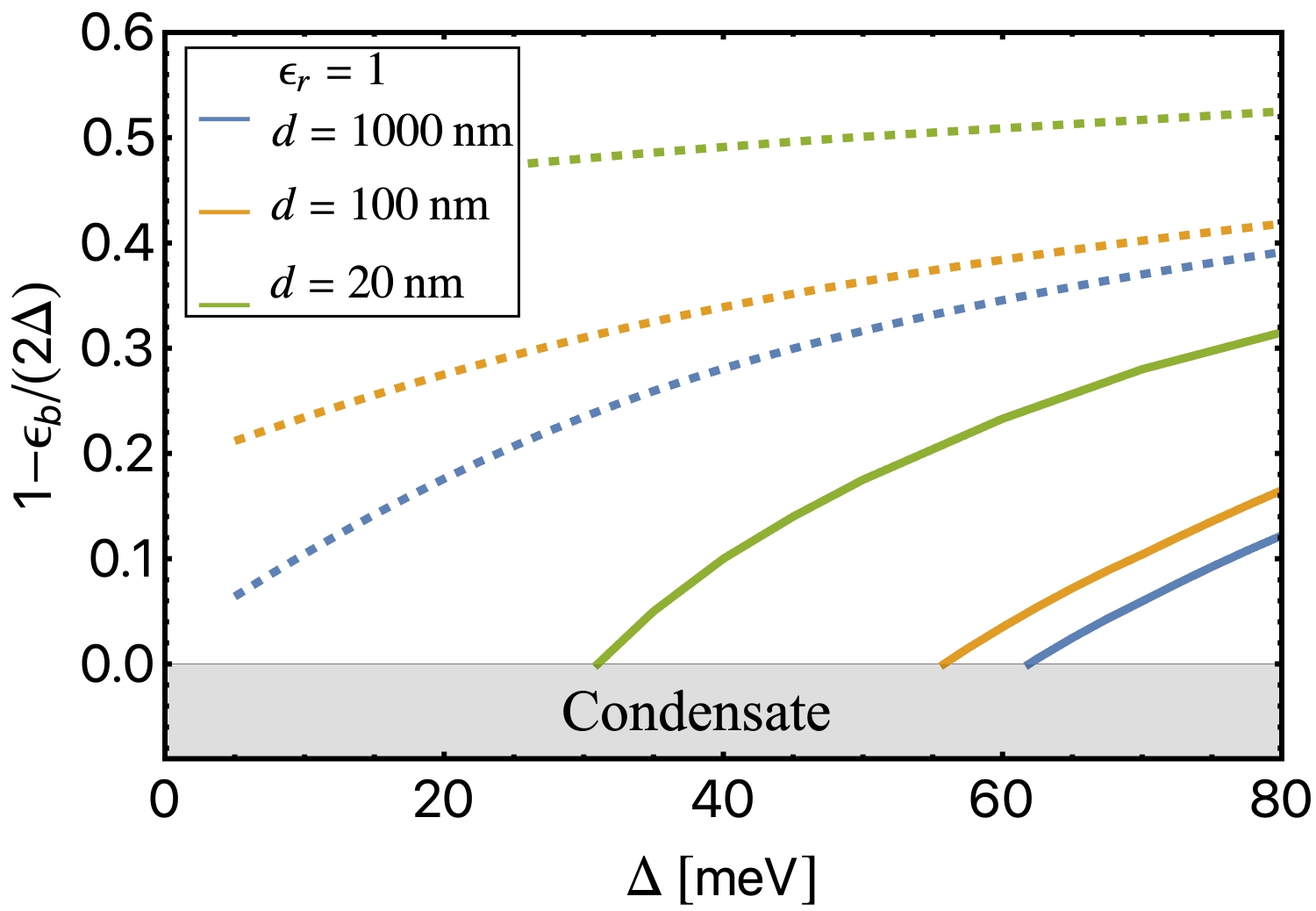}\hspace{2cm}
    \includegraphics[width=0.37505\textwidth]{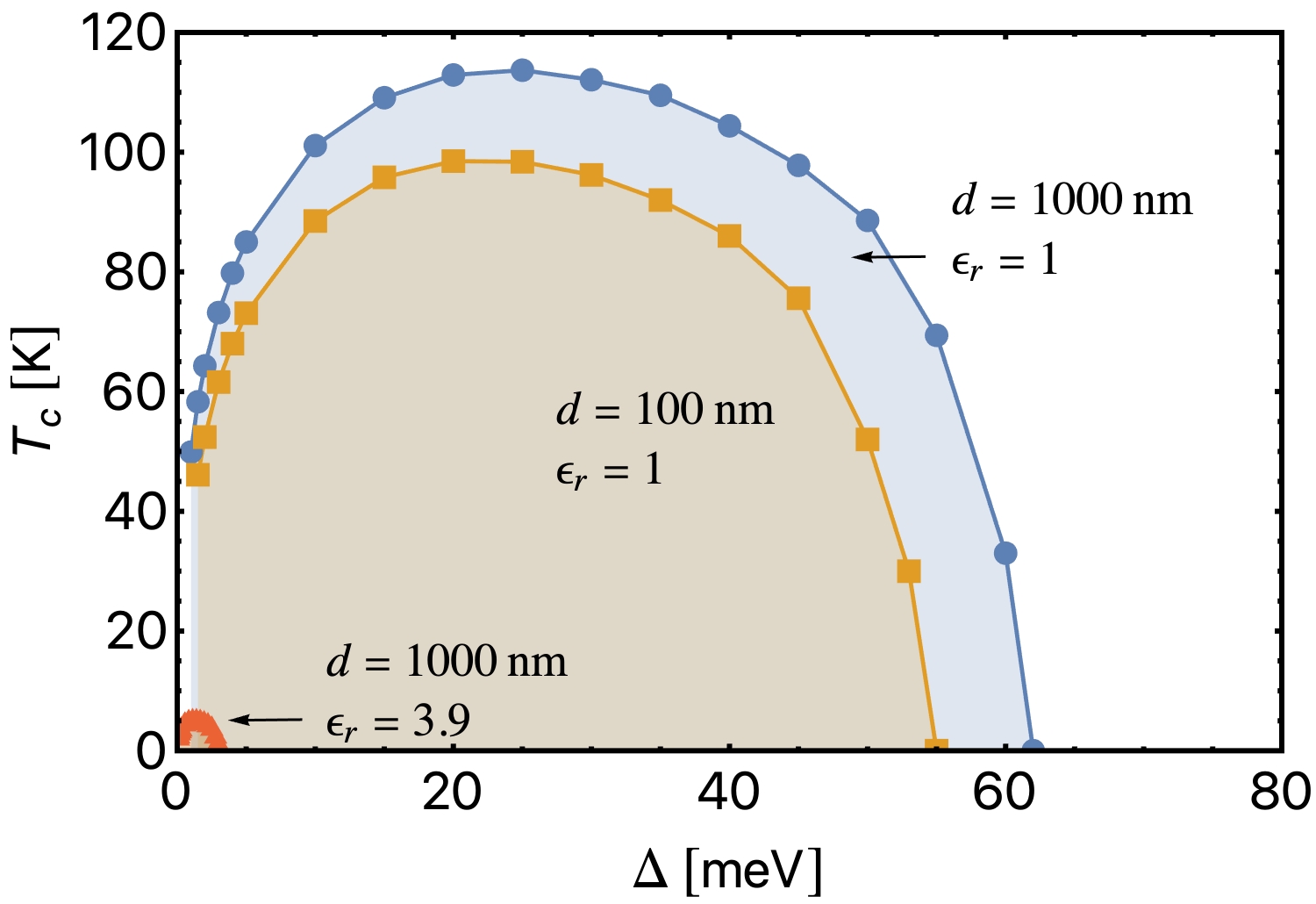}
    \caption{
 (a) Exciton binding energy at $T=0$ and three values of distance to
 gates, $d=20, 100, 1000$ nm vs the band gap, at $\epsilon_r=1$. Dashed lines correspond to LSE
 and solid lines to BSE.
(b) Superfluid domes,  for the following sets of parameters:
 \{$\epsilon_r=1$, $d=1000$nm\},
 \{$\epsilon_r=1$, $d=100$nm\},
\{$\epsilon_r=3.9$, $d=1000$nm\}.
 Points represent $T_c$ versus the
 band gap parameter $\Delta$, as computed from the BSE (\ref{BSe}).
     }\label{phase}
\end{figure*}

A frequency dependence of $V_{\bm q, i\xi}$ corresponds to retardation of the
interaction in the time-domain. For a weakly-bound exciton in an insulator, with binding energy $\epsilon_b \ll 2\Delta$,
retardation is not important \cite{ScammellSushkov2022}. However, retardation becomes essential for strongly-bound excitons $\epsilon_b \sim 2\Delta$. We point out that the condition for exciton condensation, $\epsilon_b = 2\Delta$, necessarily implies strongly-bound excitons and therefore retardation is important. To properly treat retardation in the exciton binding problem, one has to employ the Bethe-Salpeter equation (BSE) \cite{berestetskii1982quantum}
\begin{align}
\label{BSe}
&\chi_{\xi_n,\bm k}=-T \sum_m \int \frac{d^2 k'}{(2\pi)^2} \frac{V_{{\bm k-\bm k'},i(\xi_n-\xi_m)}Z_{\bm {k,k'}}^{\tau',\tau}}{(E/2-\omega_{\bm k})^2+\xi_n^2} \chi_{\xi_m,\bm k'},
\end{align}
written in terms of the amputated two-particle Green's function $\chi_{\xi_n,\bm k}$. 
Here $E=2\Delta -\epsilon_b$, $\xi_n=(2n+1)\pi T$ and  $Z_{\bm {k,k'}}^{\tau',\tau}=\langle\psi^{(-)}_{\bm {k},\tau'}|\psi^{(-)}_{\bm k',\tau'}\rangle\langle\psi^{(+)}_{\bm {k}',\tau}|\psi^{(+)}_{\bm k,\tau}\rangle$ is the vertex form factor, with $\psi^{(\pm)}_{\bm {k},\tau}$ denoting the single particle wavefunctions for conduction and valence bands of \eqref{H1}, see also \cite{Supp}. The total momentum of the electron and hole is encoded in
the valley indices, and is either zero for $\tau=\tau'$ (intravalley pairing)
or non-zero for $\tau=-\tau'$ (intervalley pairing) -- as depicted in Fig. \ref{valley}.
The form factors do not distinguish spin, yet they weakly distinguish between intra- and intervalley pairing; the implications for the condensate phase is discussed later. The interaction has an SU(2)$\times$SU(2) spin-symmetry -- correspondingly, the exciton bound state has an SO(4) spin symmetry relating the spin-singlet and triplet configurations.

We pause to mention that under an instantaneous interaction, $V_{{\bm k-\bm k'}}$, e.g. as obtained by ignoring the frequency dependence of screening, the exciton can be described
by a wave function that obeys the Lippmann-Schwinger
equation (LSE),
\begin{align}
  \label{LSE}
(E_0-2\omega_{\bm k} ) \Psi_{\bm k} =\int \frac{d^2k'}{(2\pi)^2}
         V_{{\bm k-\bm k'}}Z_{\bm {k,k'}}^{\tau',\tau}\tanh\left(\frac{\omega_{\bm k'} }{2T}\right)\Psi_{\bm k'}.
\end{align}
LSE is a linear eigenvalue problem and can be easily solved to find the eigenenergy $E_0$ and eigenfunction $\Psi_{\bm k}$; solution of (\ref{LSE}) shows that the Lippmann-Schwinger wave function
$\Psi_{\bm k}$  is well localised in the momentum space, $k \lesssim \sqrt{2m\Delta}$. This corresponds to the exciton spatial size $r\sim 1/\sqrt{2m\Delta}$. We see a hint that the wavefunction approach is problematic at $\Delta\to0$. We denote the binding energy computed from LSE as $\epsilon_b^0=2\Delta-E_0$.

{\it Phase diagram ---}
Via direct computation of Eq. \eqref{BSe}, the intervalley $s$-wave exciton is found to have the lowest energy for $\Delta>0$; we henceforth specialise to this state. 
Let us start from $T$=0. Fig. \ref{phase}a shows the exciton
binding energy vs the gap parameter $\Delta$ for $\epsilon_r=1$ and three values of
distance to the gate  $d=\{20, 100, 1000\}$nm; the solid and dashed lines are computed from the BSE (\ref{BSe}) and  LSE (\ref{LSE}), respectively, whereby only the BSE includes retardation effects. 
First, we see that retardation significantly influences the binding energy  --- one could think of retardation of the screened potential acting as an effective dynamic boson mode which is enhancing the binding of electrons and holes.
Second, the plot marks a condensation region, whereby $\epsilon_b\geq2\Delta$. At $\epsilon_r=1$, condensation occurs for gates placed beyond a critical distance $d>d_c\approx 10-15$ nm. In particular, for $d=\{20, 100, 1000\}$nm the quantum critical point is
$\Delta_c=\{31, 55, 62\}$meV. Considering instead hBN encapsulation, such that $\epsilon_r=3.9$, and taking $d=100$ nm, the critical point is reduced to $\Delta_c=3$ meV.  These results highlight the strong influence of gate and dielectric screening on the excitonic binding energy and condensation transition.

To establish the phase boundary of Fig. \ref{phase}(b), we solve $\epsilon_b(T_c)=2\Delta$. We note that for $\Delta < \Delta_c$, the two-particle problem makes sense only at or above the critical temperature, where $T_c$ is understood to be a BKT transition temperature. For $\epsilon_r=1$ and  $d=\{100, 1000\}$nm maximum
critical temperatures are significant, $T_c\approx\{100, 115\}$K, respectively. For $\epsilon_r=3.9$, $d=100$nm the superfluid dome is comparatively small.
Fig. \ref{phase}(b) shows that as $\Delta\to0$, the $T_c(\Delta)$ is rapidly decreasing. However, we stress that we do not propagate the BSE technique down to exactly $\Delta=0$; here the system becomes semi-metallic and the two-body technique employed here becomes prohibitively expensive numerically. Physically, the strong suppression of $T_c$ at small band gap $\Delta$ is due to the enhanced thermal excitation to the conduction band; the thermally excited states act as a source of metallic screening, and thereby have a significant affect on the Coulomb attraction. This feature, i.e. strongly enhanced screening, makes this problem highly non-BCS; BCS provides a simple relation between $T=0$ order parameter and $T_c$, which derives from the Pauli blocking factor (i.e. thermal occupation factors). In our case, we have both (thermal) Pauli blocking as well as thermal screening of the interaction. We find that the thermal screening plays the dominant role in melting the order, and thus we do not recover the standard BCS relation. 

Finally, we mention that within the BSE two-body formalism we arrive at the following critical scaling of $T_c$ near $\Delta\to\Delta_c^-$ (see Supplement \cite{Supp}),
\begin{align}
T_c(\Delta)\sim\Delta_c/\ln\left(\frac{g \Delta_c}{\Delta_c-\Delta}\right)
\end{align}
where $g$ is a dimensionless combination of the interaction strength and density of states, as well as other dimensionless numerical factors that we do not evaluate.

{\it Condensate phase} --- Entering the condensate phase we can no longer apply two-body techniques, and therefore do not expect to make quantitative predictions. Even so, the results below provide crucial predictions for future experimental tests of the exciton condensate. 
The discussion from here on will be based on a mean-field Hamiltonian, $H_\text{MF} = \sum_{\bm p, \tau,s}  c^\dag_{\bm p,\tau,s} \omega_{\bm p}  c_{\bm p,\tau,s} + v^\dag_{\bm p,\tau,s} (-\omega_{\bm p}) v_{\bm p,\tau,s} + \sum_{\bm p, \tau,\tau',s,s'} c^\dag_{\bm p,\tau,s} \Phi(p)_{\tau,\tau,s,s'}v_{\bm p,\tau',s'}$ +h.c., with $c^\dag_{\bm p,\tau,s} (v^\dag_{\bm p,\tau,s})$ the creation operators for conduction (valence) electrons, and with $\Phi(p)_{\tau,\tau,s,s'}$ the order parameter. Explicitly for the $s$-wave intervalley excitonic condensate, it takes the form
\begin{align}
\label{Phi}
\Phi(p)_{\tau,\tau',s,s'}=\Phi_0(p) (\tau_x)_{\tau,\tau'} e^{i\frac{(\tau-\tau')}{2}\phi} (d_\mu s_\mu i s_y)_{s,s'},
\end{align}
where $\Phi_0(p)$ is the amplitude; $\phi$ an arbitrary phase encoding the phase difference between the two distinct intervalley states, i.e. those with valley indices $(\tau,\tau')=(+-)$ and $(-+)$; $\tau_x$ is a Pauli matrix acting on valley indices; $s_\mu$ are Pauli spin matrices and $d_\mu$ are the components of a unit four-vector, with $\mu=0$ corresponding to a spin-singlet and $\mu=1,2,3$ to the components of the spin-triplet -- there is an SO(4) degeneracy of this spin ordering vector $(d_\mu)$.

 We pause to note that a spontaneous symmetry breaking in {\it unbiased} bilayer graphene has been considered \cite{Nandkishore2010}. Specifically, that work predicts a spontaneous ferroelectric polarisation perpendicular to the
plane, breaking a $\mathbb{Z}_2$ `layer' symmetry. By contrast, in the present work the applied electric bias explicitly breaks the $\mathbb{Z}_2$ symmetry and drives a single particle band gap ($2\Delta$). Our key prediction is the spontaneous breakdown of a U(1) symmetry, and hence the onset of superfluidity.
We stress that the superfluid order parameter [Eq. \eqref{Phi}] is not a ferroelectric; it does not couple linearly to an external electric field. However, it does influence the layer polarization in a measurable way and this opens a unique way to detect the superfluid quantum phase transition. We discuss these details next.

Within the ordered phase $\Delta<\Delta_c$, and at $T=0$, we appeal to the BCS/Eliashberg gap equation to estimate $\Phi_0(p)$ [for simplicity we, for now, ignore the momentum dependence, such that $\Phi_0(p)=\Phi_0$]. Near the critical point $\Delta\to\Delta_c$, and to logarithmic accuracy, the gap equation gives $1=g' \ln(\Lambda'/\sqrt{\Delta^2+|\Phi_0|^2})$ or $|\Phi_0|=\text{Re}\sqrt{\Delta_c^2-\Delta^2}$. Here $\Delta_c=\Lambda' e^{-1/g'}$, $g'\sim g$ is a dimensionless combination of the interaction and density of states and $\Lambda'\sim\Lambda$ is a UV cut-off.

The natural, measurable quantity of the system is not $\Phi_0$, but instead the layer polarization. In the absence of excitonic order, $\Phi_0=0$, the polarization is ${\cal P}_{\Delta}$ of Eq. \eqref{P}. Polarization results from the valence electrons belonging predominantly to, say, the bottom layer. Since $\Phi_0\neq0$ ultimately corresponds to removing valence electrons (bottom layer) and enhancing conduction electrons (top layer), one expects $\Phi_0\neq0$ to reduce the ground state polarization. 
The polarization in the condensate phase is then simply ${\cal P}_{\Delta,\Phi_0}=\frac{2m\Delta}{\pi}\ln\left(\Lambda/\sqrt{\Delta^2+|\Phi_0|^2}\right)$. We see that $\Phi_0\neq0$ indeed reduces the layer polarization, with leading correction quadratic in $\Phi_0$. 
Further, we appeal to the gap equation solution to arrive a key prediction, 
\begin{align}
{\cal P}_{\Delta,\Phi_0} - {\cal P}_{\Delta} = -\frac{m |\Phi_0|^2}{\pi\Delta} \xrightarrow[\Delta\to\Delta_c^-]{} -\frac{2m (\Delta_c-\Delta)}{\pi}.
\label{deltaP}
\end{align}
This expression shows that there is a linear in $\Delta$ (and hence in external electric field $E$) reduction of the layer polarization in the vicinity of the critical point.  We anticipate the polarization being readily measured via quantum capacitance, see e.g. \cite{YangQC1997, DuQC2017, SaberiQC2020}, which would provide a direct experimental test of the exciton condensation transition. 

\begin{figure}[t!]
  \includegraphics[width=0.375\textwidth]{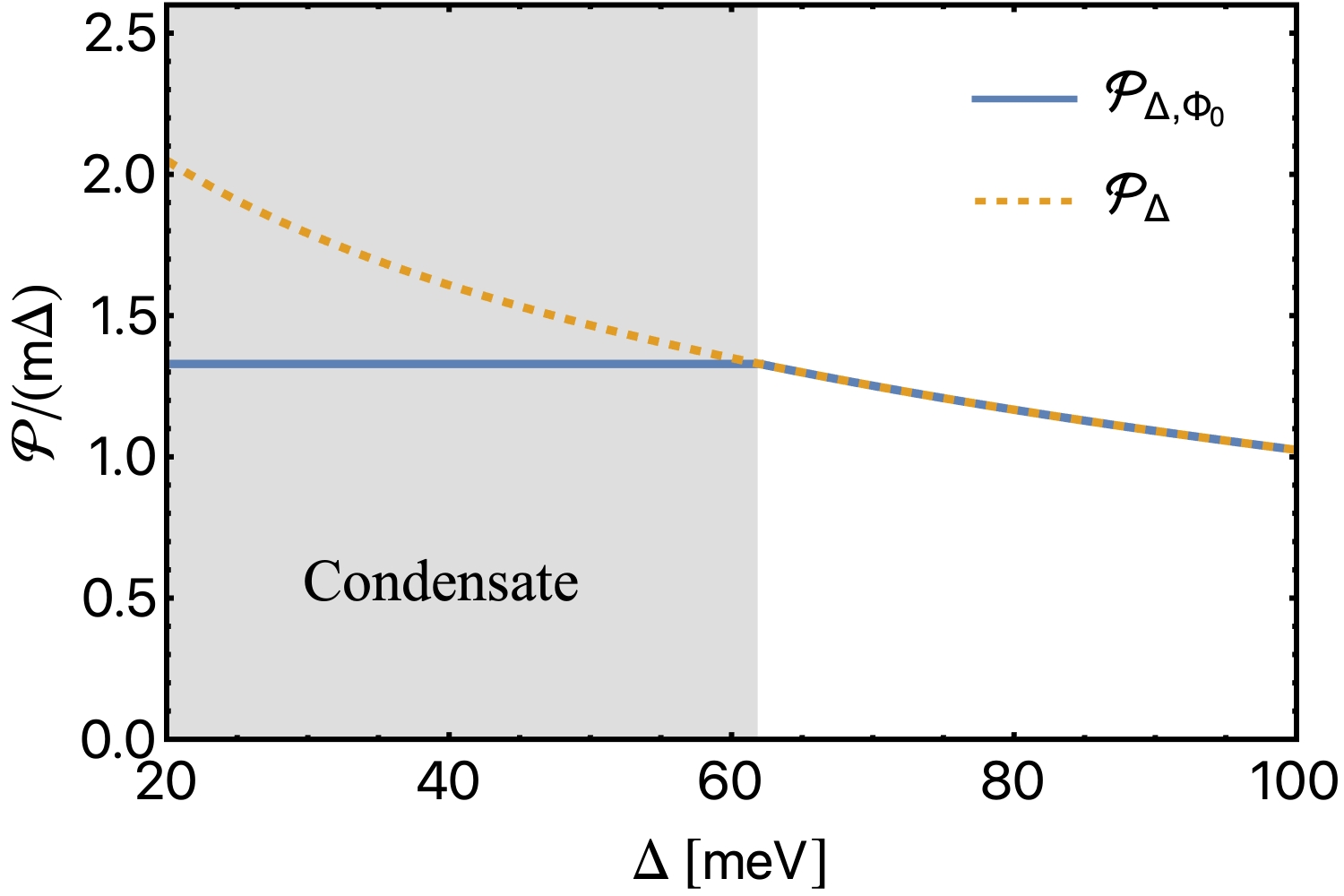}
    \caption{Dimensionless layer polarization, as a function of band gap parameter $\Delta$, with and without the condensate -- corresponding to ${\cal P}_{\Delta,\Phi}$ and ${\cal P}_{\Delta}$ [Eq. \eqref{P}] , respectively. Here we use the meanfield expression $|\Phi_0|=\text{Re}\sqrt{\Delta_c^2-\Delta^2}$, at $T=0$, with $\Delta_c=62$ meV corresponding to system parameters $\epsilon_r=1, d=1000$ nm. Due to the onset of the condensate $\Phi$, a cusp is seen in reduced layer polarization, ${\cal P}_{\Delta,\Phi}/(m\Delta)$, at the critical point $\Delta_c$.}
    \label{layerPol}
\end{figure}

Finally, we turn to the in-plane real-space structure of the condensate.  
The intervalley ordering [Eq. \eqref{Phi}] implies that, in real-space, the condensate exhibits a pair density wave pattern, denoting $\tilde{\Phi}_{s,s'}(\bm r, \bm r')=\tilde{\Phi}(\bm r, \bm r')(d_\mu s_\mu i s_y)_{s,s'}$, 
\begin{align}
\label{RealSpace}
 \tilde{\Phi}(\bm r, \bm r')&= \tilde{\Phi}_0(\bm r - \bm r') \cos(\bm K\cdot(\bm r + \bm r')+\phi) 
\end{align}
with $\Phi_0(\bm r - \bm r')$ a complex amplitude (derived in the Supplement \cite{Supp}). Fluctuations of $\phi$ correspond to gapless sliding modes of the pair density wave. The pair density wave is spatially modulated with a periodicity of three unit cells.  In the context of superconductivity, such a density wave pattern has been analysed at length in Ref. \cite{RoyKekule2010} and has been linked to higher-order topology in Refs. \cite{LiIngham2020, LiGeier2021} -- we leave the question of higher topology of the mean-field order Eq. \eqref{RealSpace} for future work.

\

{\it Discussion} --- We presented a description for the excitonic bound states in biased BLG, and their subsequent condensation. Owing to biased BLG being a band insulator, we were able to apply two-body techniques. This is contrasted to exciton formation and condensation from a semi-metallic ground state, whereby many-body techniques are necessary, and resultantly uncontrolled approximations are often required.

We examined suspended biased BLG, with gates $d>d_c\approx 10-15$ nm. In this regime/setup, we show that excitons condense for $\Delta<\Delta_c$, Fig. \ref{phase}(a). This is our primary theoretical finding. For distance to gates $d=100$nm (1000nm) the maximum critical temperature is predicted to be $100$K ($115$K), and is optimised at non-zero bias, Fig. \ref{phase}(b). Entering the BEC phase, our two-body techniques are no longer quantitatively accurate, however, we are able to establish several qualitative features of the condensate. Experimentally distinguishing between the excitonic condensate/insulator and the band insulator is a challenging task. To this end, we found that the excitons influence the macroscopic dipole moment [c.f. \eqref{deltaP}], and proposed a smoking gun test for exciton condensation -- quantum capacitance measurements of the dipole moment as a function of the bias field, which tunes the band insulator to exciton condensate phase transition [c.f. Fig. \ref{layerPol}]. 
We hope to entice direct experimental searches for the excitonic condensate in biased bilayer graphene. Future theoretical work may consider other candidate materials using the two-body approach applied here.

\begin{acknowledgements}
We thank Dmitry Efimkon, Alex Hamilton, Mike Zhitomirsky and J\"org Schmalian for enlightening conversations. We acknowledge funding support from the Australian Research Council Centre of Excellence in Future Low-Energy Electronics Technology (FLEET) (CE170100039).
\end{acknowledgements}


%

\clearpage

\begin{widetext}

\section*{Supplemental Material}

\newcommand{\beginsupplement}{%
        \setcounter{table}{0}
        \renewcommand{\thetable}{S\arabic{table}}%
        \setcounter{figure}{0}
        \renewcommand{\thefigure}{S\arabic{figure}}%
                \setcounter{equation}{0}
        \renewcommand{\theequation}{S\arabic{equation}}%
     }

\beginsupplement

\renewcommand{\theequation}{S\arabic{equation}}

\section{Hamiltonian and wavefunctions}
The Hamiltonian and wavefunctions are (in zero field)
\begin{align}
\notag H_\tau&=-\frac{(\tau p_x \mp i p_y)^2}{2m}\sigma_\pm + \Delta\sigma_z\\
\notag |\psi^{(+)}_{\bm p,\tau}\rangle&=\frac{1}{\sqrt{\frac{(\varepsilon_{\bm p}-\Delta)^2}{p^4/(4m^2)}+1}}\begin{pmatrix} -1 \\ \frac{\varepsilon_{\bm p}-\Delta}{p^2/(2m)}e^{2i\tau \theta_{\bm p}} \end{pmatrix},\\
\notag |\psi^{(-)}_{\bm p,\tau}\rangle= {\cal P} |\psi^{(+)}_{\bm p,\tau}\rangle&= \frac{1}{\sqrt{\frac{(\varepsilon_{\bm p}-\Delta)^2}{p^4/(4m^2)}+1}}\begin{pmatrix} \frac{\varepsilon_{\bm p}-\Delta}{p^2/(2m)}e^{-2i\tau \theta_{\bm p}} \\ 1 \end{pmatrix},\\
{\cal P}&=i\sigma_y {\cal C}.
\end{align}  
The operator ${\cal P}$ generates the particle-hole transformation, which is an (anti)symmetry of this Hamiltonian. Defining $\psi_-$ in this way correctly generates the phase/winding factors. 

\section{Polarization Operator}\label{polop_supp}
The expression for the polarization generically takes the form,
\begin{align}
         \Pi({\bm q},i\xi,T)&=4 \sum_{\mu,\nu=\pm}  \int \frac{d^2 p}{(2\pi)^2} \frac{\left(f_{\epsilon^\mu_{\bm p}}-f_{\epsilon^\nu_{\bm {p+q}}}\right)}
    {i\xi + \epsilon^\mu_{\bm p}-\epsilon^\nu_{\bm {p+q}}}F^{\mu\nu}_{\bm {p,p+q}}.
\end{align}
Here the form factors are $F^{\mu\nu}_{\bm {p,p+q}}=|\langle\psi^{(\mu)}_{\bm {p+q},\tau}|\psi^{(\nu)}_{\bm p,\tau}
\rangle|^2$, with $\mu,\nu=\pm$.
Using that $-\epsilon^-_{\bm p}=\epsilon^+_{\bm p}\equiv \epsilon_{\bm p}$ and $F^{--}_{\bm {p,p+q}}=F^{++}_{\bm {p,p+q}}$, $F^{-+}_{\bm {p,p+q}}=F^{+-}_{\bm {p,p+q}}$, we get 
\begin{align}
  \label{polT}
         \Pi({\bm q},i\xi,T)&=8 \int \frac{d^2 p}{(2\pi)^2}\left[ \frac{(\epsilon_{\bm p}-\epsilon_{\bm {p+q}})\left(f_{\epsilon_{\bm p}}-f_{\epsilon_{\bm {p+q}}}\right)}
    {\xi^2 + (\epsilon_{\bm p}-\epsilon_{\bm {p+q}})^2}F^{++}_{\bm {p,p+q}} + \frac{(\epsilon_{\bm p}+\epsilon_{\bm {p+q}})\left(f_{\epsilon_{\bm p}}+f_{\epsilon_{\bm {p+q}}}-1\right)}
    {\xi^2 + (\epsilon_{\bm p}+\epsilon_{\bm {p+q}})^2}F^{+-}_{\bm {p,p+q}} \right].
\end{align}

\section{Vertex Form Factors}
The Coulomb interaction is taken to be
\begin{align}
H_{int} &=\sum_{\bm p_1,\bm p_2,\bm p_3}\sum_{\tau,\tau'}V(\bm p_2-\bm p_1)\Psi^\dag_{\bm p_1,\tau} \Psi^\dag_{\bm p_2,\tau'} \Psi_{\bm p_3,\tau'} \Psi_{\bm p_1+\bm p_2 -\bm p_3,\tau},
\end{align}
which neglects valley-exchange. 
We transform to band basis, 
\begin{align}
\notag \Psi_{\bm p,\tau} &= {\cal U}_{\bm p,\tau}\begin{pmatrix} c_{\bm p,\tau}\\ v_{\bm p,\tau} \end{pmatrix},\ {\cal U}_{\bm p,\tau}=(|\psi^{(+)}_{\bm p,\tau}\rangle, |\psi^{(-)}_{\bm p,\tau}\rangle),
\end{align}
with  $c_{\bm p,\tau}$ and 
$v_{\bm p,\tau}$ the destruction operators for conduction and valence electrons. 
Restricting consideration to the exciton channel, denoted $H_{int}^X$, the interaction becomes,
\begin{align}
H_{int}^X &=\sum_{\bm p_1,\bm p_2}\sum_{\tau,\tau'}V(\bm p_2-\bm p_1)c^\dag_{\bm p_1,\tau} c_{\bm p_2,\tau} v^\dag_{\bm p_2,\tau'} v_{\bm p_1,\tau'} \langle\psi^{(-)}_{\bm {p}_2,\tau'}|\psi^{(-)}_{\bm {p}_1,\tau'}\rangle\langle\psi^{(+)}_{\bm {p}_1,\tau}|\psi^{(+)}_{\bm p_2,\tau}\rangle.
\end{align}
In the main text we denote the from factor $Z_{\bm p_1,\bm p_2}^{\tau',\tau}=\langle\psi^{(-)}_{\bm {p}_2,\tau'}|\psi^{(-)}_{\bm {p}_1,\tau'}\rangle\langle\psi^{(+)}_{\bm {p}_1,\tau}|\psi^{(+)}_{\bm p_2,\tau}\rangle$.

\section{Critical temperature scaling}\label{s:Tc}
Setting $E=0$, the $T=0$ the Lippmann-Schwinger equation gives the condition for $\Delta_c$
\begin{align}
2\sqrt{\frac{p^4}{4m^2}+\Delta_c^2}\Psi_{\bm p}&=\int \frac{d^2p'}{(2\pi)^2} V^0_{\bm p,\bm p'}\Psi_{\bm p'}.
\end{align}
At $T>0$, the condition $E=0$ requires $\Delta<\Delta_c$, and is given by
\begin{align}
2\sqrt{\frac{p^4}{4m^2}+\Delta_c^2}\Psi_{\bm p}&=\int \frac{d^2p'}{(2\pi)^2} \left[V^0_{\bm p,\bm p'} + \delta V_{\bm p,\bm p'} e^{-\frac{\Delta}{T_c}}\right]\Psi_{\bm p'}.
\end{align}
Using that the characteristic momentum is $p=\sqrt{2m\Delta}$, we combine these two equations as
\begin{align}
\notag \Delta_c &= \nu_0,\\
\notag \Delta &=\left(\nu_0- \nu_1 e^{-\frac{\Delta}{T_c}}\right),\\
(\Delta_c-\Delta)&=\nu_1 e^{-\frac{\Delta}{T_c}}, \to T_c=\Delta/\ln\left(\frac{\nu_1}{\Delta_c-\Delta}\right).
\end{align}

\section{Real-space order parameter}
In this appendix we derive the effective lattice model for the dominant excitonic order parameter. 
Introducing the real-space creation operators for the conduction and valence bands via,
\begin{align}
\notag c_{\bm p,\tau,s}^\dag&=\sum_{\bm r} \phi^{(+)}_{\bm p,\tau}(\bm r) \tilde{c}^\dag_{\bm r,s},\\
v_{\bm p,\tau,s}^\dag&=\sum_{\bm r} \phi^{(-)}_{\bm p,\tau}(\bm r) \tilde{v}^\dag_{\bm r,s}.
\end{align}
With $\bm r$ spanning the real-space lattice sites, and comprising two sublattices, $\sigma=A_1,B_2$, and $\{a(\bm r\in A)=1, a(\bm r\in B)=0\}$ and $\{b(\bm r\in A)=0, b(\bm r\in B)=1\}$.
Here,
\begin{align}
\notag \phi^{(+)}_{\bm p,\tau}(\bm r)&=e^{i(\bm p + \tau \bm K)}\left(\alpha^{(+)}_{\tau,\bm k} a(\bm r) + \beta^{(+)}_{\tau,\bm k} b(\bm r)\right),\\
\phi^{(-)}_{\bm p,\tau}(\bm r)&=e^{i(\bm p + \tau \bm K)}\left(\alpha^{(-)}_{\tau,\bm k}a(\bm r) + \beta^{(-)}_{\tau,\bm k} b(\bm r)\right),
\end{align}
with definitions,
\begin{align}
\notag \alpha^{(+)}_{\tau,\bm p}&=-\cos\gamma_{\bm p} = \frac{-1}{\sqrt{\frac{(\varepsilon_{\bm p}-\Delta)^2}{p^4/(4m^2)}+1}},\\
\notag \beta^{(+)}_{\tau,\bm p}&=\sin\gamma_{\bm p}e^{2i\tau \theta_{\bm p}} = \frac{\varepsilon_{\bm p}-\Delta}{p^2/(2m)}\frac{e^{2i\tau \theta_{\bm p}}}{\sqrt{\frac{(\varepsilon_{\bm p}-\Delta)^2}{p^4/(4m^2)}+1}},\\
\notag \alpha^{(-)}_{\tau,\bm p}&=\sin\gamma_{\bm p}e^{-2i\tau \theta_{\bm p}}= \frac{\varepsilon_{\bm p}-\Delta}{p^2/(2m)}\frac{e^{-2i\tau \theta_{\bm p}}}{\sqrt{\frac{(\varepsilon_{\bm p}-\Delta)^2}{p^4/(4m^2)}+1}},\\ 
\beta^{(-)}_{\tau,\bm p}&=\cos\gamma_{\bm p}  = \frac{1}{\sqrt{\frac{(\varepsilon_{\bm p}-\Delta)^2}{p^4/(4m^2)}+1}}.
\end{align}

The mean-field Hamiltonian becomes,
\begin{align}
H_\Phi&=\sum_{\bm p,\tau,\tau',s,s'} c_{\bm p,\tau,s}^\dag \Phi(p)_{\tau,\tau',s,s'} v_{\bm p,\tau',s'} = \sum_{\bm p,\tau,\tau',s,s'} \sum_{\bm r,\bm r'} \phi^{(+)}_{\bm p,\tau}(\bm r) (\phi^{(-)}_{\bm p,\tau'})^*(\bm r') \Phi(p)_{\tau,\tau',s,s'}  \tilde{c}^\dag_{\bm r,s} \tilde{v}_{\bm r',s'}+ \text{h.c.}\ . 
\end{align}
We consider intervalley pairing in the $s$-wave channel and with arbitrary spin ordering (singlet or triplet). There is both a symmetric and an anti-symmetric combination of the valleys,
\begin{align}
\Phi(p)_{\tau,\tau',s,s'}&=
\begin{cases} \Phi_0(p) (\tau_x)_{\tau,\tau'}(d_\mu s_\mu)_{s,s'} \\ \Phi_0(p) (i\tau_y)_{\tau,\tau'} (d_\mu s_\mu)_{s,s'}.
\end{cases}
\end{align}
In fact, there is a U(1) rotational symmetry that connects these distinct valley structures -- we more compactly write
\begin{align}
\Phi(p)_{\tau,\tau',s,s'}&=\Phi_0(p) (\tau_x)_{\tau,\tau'} e^{i\frac{(\tau-\tau')}{2}\phi} (d_\mu s_\mu)_{s,s'}
\end{align}

Let us consider the case that $\bm r\in A$ and $\bm r'\in B$, 
\begin{align}
\notag H^{AB}_\Phi&=\sum_{\bm p,\tau,s,s'} \sum_{\bm r\in A,\bm r'\in B} \left(e^{i(\bm p + \tau \bm K)\cdot\bm r}\alpha^{(+)}_{\tau,\bm p} a(\bm r)\right) \left(e^{-i(\bm p - \tau \bm K)\cdot\bm r'}\beta^{(-)}_{-\tau,\bm p} b(\bm r')\right) \Phi_0(p) e^{i\tau\phi}\tilde{c}^\dag_{\bm r,s} \tilde{v}_{\bm r',s'} (d_\mu s_\mu)_{s,s'} + \text{h.c.}\ \\
\notag &=\sum_{\bm p,\tau,s,s'} \sum_{\bm r\in A,\bm r'\in B} \left[e^{i\bm p\cdot(\bm r - \bm r')}e^{i\tau \bm K\cdot(\bm r + \bm r')} \alpha^{(+)}_{\tau,\bm p} \beta^{(-)}_{-\tau,\bm p}  \Phi_0(p)e^{i\tau\phi} \right]  \tilde{c}^\dag_{\bm r,s} \tilde{v}_{\bm r',s'} (d_\mu s_\mu)_{s,s'}+ \text{h.c.}\ ,\\
\notag &= \sum_{\bm r\in A,\bm r'\in B}\sum_{\bm p} \left[2 e^{i\bm p\cdot(\bm r - \bm r')} (-\cos^2\gamma_{\bm p}) \Phi_0(p)\right]  \left[\cos(\bm K\cdot(\bm r + \bm r')+\phi)\right] \sum_{s,s'} \tilde{c}^\dag_{\bm r,s} \tilde{v}_{\bm r',s'} (d_\mu s_\mu)_{s,s'}+ \text{h.c.}\ ,\\
&= \sum_{\bm r\in A,\bm r'\in B}\Gamma^{AB}(\bm r - \bm r') \left[\cos(\bm K\cdot(\bm r + \bm r')+\phi)\right] \sum_{s,s'} \tilde{c}^\dag_{\bm r,s} \tilde{v}_{\bm r',s'} (d_\mu s_\mu)_{s,s'} + \text{h.c.}\ .
\end{align}
We have defined the function, 
\begin{align}
\Gamma^{AB}(\bm r - \bm r')&\equiv\sum_{\bm p} \left[2 e^{i\bm p\cdot(\bm r - \bm r')} (-\cos^2\gamma_{\bm p}) \Phi_0(p)\right],
\end{align}
which determines the real-space amplitude and can be directly evaluated numerically. 

Performing the same procedure for all combinations of $\bm r, \bm r'$ in $\{A_1, B_2\}$, we arrive at the expression, 
\begin{align}
H_\Phi&= \sum_{\bm r,\bm r}\Gamma(\bm r - \bm r') \left[\cos(\bm K\cdot(\bm r + \bm r')+\phi)\right] \sum_{s,s'} \tilde{c}^\dag_{\bm r,s} \tilde{v}_{\bm r',s'} (d_\mu s_\mu)_{s,s'} + \text{h.c.},
\end{align}
with amplitude function,
\begin{align}
\Gamma(\bm r - \bm r')&=
\begin{cases} \Gamma^{AA}(\bm r - \bm r')=\sum_{\bm p} \left[2 e^{i\bm p\cdot(\bm r - \bm r')} (-\cos\gamma_{\bm p}\sin\gamma_{\bm p} e^{\pm2i\tau\theta_{\bm p}}) \Phi_0(p)\right], \ \text{for} \ \{\bm r\in A_1,\bm r'\in A_1\}  \\ \Gamma^{BB}(\bm r - \bm r')=\sum_{\bm p} \left[2 e^{i\bm p\cdot(\bm r - \bm r')} (\cos\gamma_{\bm p}\sin\gamma_{\bm p} e^{\pm2i\tau\theta_{\bm p}}) \Phi_0(p)\right], \ \text{for} \ \{\bm r\in B_2,\bm r'\in B_2\} \\ \Gamma^{AB}(\bm r - \bm r')=\sum_{\bm p} \left[2 e^{i\bm p\cdot(\bm r - \bm r')} (-\cos^2\gamma_{\bm p}) \Phi_0(p)\right], \ \text{for} \ \{\bm r\in A_1,\bm r'\in B_2\} \\ \Gamma^{BA}(\bm r - \bm r') = \sum_{\bm p} \left[2 e^{i\bm p\cdot(\bm r - \bm r')} (\sin^2\gamma_{\bm p} e^{\pm4i\theta_{\bm p}}) \Phi_0(p)\right], \ \text{for} \ \{\bm r\in B_2,\bm r'\in A_1\} .
\end{cases}
\end{align}

\end{widetext}

\end{document}